\documentclass[12pt]{article}
\def\be{\begin{equation}}
\def\ee{\end{equation}}
\def\ba{\begin{array}{c}}
\def\ea{\end{array}}

\def\ben{\[ }
\def\een{\] }
\usepackage{epsfig}

\begin{document}

\title{
Comment on `Solution of the Dirac equation for the Woods-Saxon
potential with spin and pseudospin symmetry' [J. Y. Guo and Z-Q.
Sheng, Phys. Lett. A 338 (2005) 90] }
\author{
H. B\'\i la, V. Jakubsk\'y and  M. Znojil\\
{\small {\sl \'{U}stav jadern\'e fyziky AV \v{C}R, 250 68
\v{R}e\v{z}, Czech Republic$\thanks{Emails: hynek.bila@ujf.cas.cz,
\ jakub@ujf.cas.cz, \ znojil@ujf.cas.cz}$ }}}

\thispagestyle{empty}

\begin{titlepage}
%\date{\today}

\maketitle

\thispagestyle{empty}

%========================================================================
%\newpage

\section*{Abstract}

The bound-state method employed by Guo and Sheng ({\it loc.~cit.})
is shown inadequate since only one of their solutions remains
compatible, in the spin-symmetric low-mass regime, with the
physical boundary conditions. We clarify the problem and construct
a new, correct solution in the pseudospin-symmetric regime.

\vspace{0.5cm}

\noindent PACS number(s): 03.65.Pm, 03.65.Ge, 02.30.Gp

\noindent Keywords: Schr\"{o}dinger equation, Dirac equation,
Woods-Saxon potential

\end{titlepage}

%========================================================================
\newpage
%\section{Introduction and summary}

 \noindent
Recent letter \cite{cinani} paid attention to the radial Dirac
$s-$wave bound-state problem considered in the two different
dynamical regimes. This problem has been reduced to the solution
of the respective Schr\"{o}dinger-like bound-state equations
 \be
 -\frac{d^2}{dr^2}
 \,F
 (r) + (M+E-C)\,[M-E+\Sigma(r)]
 \,F
 (r)=0, \ \ \ \ \ \  F(0)=F(\infty)=0, \ \ \
   \label{5AHO}
 \ee
(cf. eqs.~(5) or (9) in \cite{cinani}) and
 \be
 -\frac{d^2}{dr^2}\,G
 (r) + (M-E+C)\,[M+E-\Delta(r)]
 \,G
 (r)=0, \ \ \ \ \ \  G(0)=G(\infty)=0,
   \label{6AHO}
 \ee
(cf. eqs.~(6) or (29) in \cite{cinani}) where $M$ denotes the
mass, $E$ is the bound-state energy, $C$ is a parameter and the
functions $\Sigma(r)$ and $\Delta(r)$ represent certain
phenomenological external potentials which have been chosen in the
elementary Woods-Saxon form,
 \be
 \label{eqWS}
 \Sigma(r)=-\frac{\Sigma_{0}}{1+\exp \left ({\frac{r-R_{}}{a}}\right )}\,,
\ \ \ \ \ \
 \Delta(r)=-\frac{\Delta_{0}}{1+\exp \left ({\frac{r-R_{}}{a}}\right
 )}\,.
  \end{equation}
In the non-relativistic context, the solution of a very similar
differential equation has been described in fair detail in
ref.~\cite{Fluegge}. Unfortunately, the existence of certain
specific {\em physical} features of the relativistic
eqs.~(\ref{5AHO}) and (\ref{6AHO}) forced one of us to imagine
\cite{mr} that the constructions of ref.~\cite{cinani} should be
re-examined.

In a preparatory step the authors of ref.~\cite{cinani} replaced
the combinations of constants appearing in eqs.~(\ref{5AHO}),
(\ref{6AHO}) and (\ref{eqWS}) by the life-simplifying
abbreviations
 \ben
 \mu^2_{(i)}=(M-E)(M+E-C)a^2, \ \ \ \ \ \
 \mu^2_{(ii)}=(M+E)(M-E+C)a^2,
 \een
 \ben
 \nu^2_{(i)}=(M+E-C)\Sigma_0a^2, \ \ \ \ \ \
 \nu^2_{(ii)}=-(M-E+C)\Delta_0a^2.
 \een
This converts both eqs.~(\ref{5AHO}) and (\ref{6AHO}) to the same
eigenvalue problem,
 \be
 -a^2\,\frac{d^2}{dr^2}
 \,y
 (r) + \mu^2
 \,y
 (r)
 - \frac{\nu^2}{1+\exp[(r-R)/a]}\,y(r)
 =0, \ \ \ \ \ \  y(0)=y(\infty)=0.
   \label{56AHO}
 \ee
The elementary change of variables
 \ben
 r \to x=1/(1+\exp
[(r-R)/a])
 \een
reduces eq.~(\ref{56AHO}) to the Gauss's hypergeometric
differential equation (compare eq.~(64.3) in \cite{Fluegge} with
the same formulae preceding eqs.~(9) and (29) in \cite{cinani}).
One arrives at the exact wave functions sampled, say, by the
unnumbered equation preceding eq.~(64.10) in the non-relativistic
construction \cite{Fluegge}. Its relativistic analogue
 \ben
 \psi(x) = N \,x^\mu
 \left [(1-x)^\delta\,f(\mu,\delta)\,_2F_1(\mu+\delta,
 \mu+\delta+1, 1+2\delta; 1-x) \right .
 \een
 \be
 \ \ \ \ \ \ + \left .
 (1-x)^{-\delta}\,f(\mu,-\delta)\,_2F_1(\mu-\delta,
 \mu-\delta+1, 1-2\delta; 1-x)\right ]\,
 \label{zakl11}
 \ee
contains a certain ratio $f(\mu,\delta)$ of products of pairs of
$\Gamma-$functions given in full detail in eq.~~(13) of
\cite{cinani}. In this notation, the boundary condition in the
origin becomes reduced to the single and exact transcendental
equation
 \be
 \psi
 \left (
 \frac{1}{1+e^{{-R_{}}/{a}}}
 \right )=0\,.
 \label{mainequation}
 \ee
Usually \cite{Fluegge}, the purely numerical {\em exact}
quantization condition (\ref{mainequation}) is being replaced by
its {\em approximate} asymptotic simplification
 \be
 \psi
 \left (1-e^{{-R_{}}/{a}}
 \right )\cdot \left [1 + {\cal O}
 \left ( e^{{-R_{}}/{a}}\right )
 \right ]
 =0\,, \ \ \ \ \ \ \ \ \ \ R/a \gg 1\,.
 \label{apprequation}
 \ee
The authors of ref. \cite{cinani} did the same [cf. their four
main quantization conditions (17), (22), (32) and (37)]. In this
sense, the slightly misleading first line of their abstract
(saying that their ``\ldots Dirac equation is solved exactly
\ldots") should rather read ``\ldots Dirac equation is solved
approximatively \ldots".

(i) In the first, spin-symmetric Dirac's dynamical regime the
authors of ref. \cite{cinani} had to separate the solution of the
ordinary differential Dirac bound-state problem (\ref{5AHO}) in
two subcases. Incidentally, the first subcase [to be numbered or
indexed as (i.a) in what follows] provides the only case where the
result of ref.~\cite{cinani} is correct. We shall show that their
construction {\em fails} to incorporate the boundary conditions
properly in {\em all} the other cases.

(i.a) Assuming that $\mu^2_{(i)}=\mu^2>0$ and returning to eq.~(9)
of \cite{cinani} in the first subcase one deduces that
$\nu^2_{(i)}=\nu^2>0$ and requires that the potential is
sufficiently attractive, i.e., that $\nu^2-\mu^2 \equiv
\gamma^2>0$. In such a case, strictly speaking, the extraction of
the {\em exact} spectrum from its implicit definition
(\ref{mainequation}) represents a straightforward, albeit {\em
purely numerical} task. Unpleasant due to the well known slowness
of the convergence of the infinite hypergeometric series
 \ben
 _2\!F_1(a,b,c;z) = 1 +\frac{ab}{c}z
 + \frac{a(a+1)b(b+1)}{2c(c+1)}z^2 +
 \ldots
 \een
in $\psi(x)$. Fortunately, in eq.~(\ref{zakl11}) the value of the
argument $z$ remains exponentially small whenever $R/a \gg 1$.
This means that in eq.~(\ref{apprequation}) re-written in its
fully explicit form
 \ben
 \left [{e^{-R/a}}
 /\left ( {1+e^{-R/a}}\right )\right]^{2\delta}
 f(\mu,\delta)
 \cdot\, _2\!F_1\left(\mu+\delta,
 \mu+\delta+1, 1+2\delta; {e^{-R/a}}
 /\left ({1+e^{-R/a}}\right )\right)=
 \een
 \ben
  =-f(\mu,-\delta)\,\cdot \,_2\!F_1\left(\mu-\delta,
 \mu-\delta+1, 1-2\delta; {e^{-R/a}}
 /\left ({1+e^{-R/a}}\right )\right)
 % \label{zakl12}
 \een
the approximation $_2\!F_1(a,b,c;z) \to 1$ will lead to very good
precision. The authors of ref. \cite{cinani} paralleled the
non-relativistic exercise of ref. \cite{Fluegge} and obtained the
definitions (20) and (21) of the two components $F_n(r)$ and
$G_n(r)$ of the related Dirac's bound-state wave functions. Their
parallel implicit formula (19) for energies $E_n$
 \be
 {\rm arg}\,\Gamma(2{\rm i}\gamma)-
 {\rm arg}\,\Gamma(\mu +{\rm i}\gamma)
 - {\rm arctan} (\gamma/\mu) +\gamma R/a = (n+1/2)\pi\,
 \label{origo}
 \ee
was simply based on the approximation (\ref{apprequation}).

Let us note here that the next-order truncation $_2\!F_1(a,b,c;z)
\to 1+abz/c$ clearly and manifestly confirms the expectations as
it leads to an elementary explicit incorporation of the next-order
corrections via transition from eq.~(\ref{origo}) to another
closed formula,
 \be
 {\rm arg}\,\Gamma(2{\rm i}\gamma)-
 {\rm arg}\,\Gamma(\mu +{\rm i}\gamma)
 - {\rm arctan} \frac{\gamma}{\mu} +\gamma \frac{R}{a}
 +
 \left(\!
 \mu +\gamma + \frac{\mu^2+\gamma^2}{1+4\gamma^2}
 \right )\exp \left (\!\!-\frac{R}{a} \right )
  \!=\! \left (\!n+\frac{1}{2}\right )\pi.
 \label{origotr}
 \ee
This improvement illustrates, persuasively enough, the exponential
smallness of the higher-order corrections under the relativistic
kinematics.

(i.b) In an alternative range of parameters one admits that both
the values of $\mu$ and $\nu$ may be imaginary, i.e.,
$\mu^2_{(i)}=\mu^2\leq 0$ and $\nu^2_{(i)}=\nu^2\leq 0$ while
$\mu^2-\nu^2 \equiv \delta^2\geq 0$ (cf.~\cite{cinani}). It is
fairly elementary to notice that at the very large $r \gg 1$ we
only have to deal with the significantly simplified
constant-coefficient form of the original radial Dirac equation
(\ref{5AHO}),
 \be
 -{a^2}\,\frac{d^2}{dr^2}
 \,F
 (r) + {\mu^2_{(i)}}
 \,F
 (r)=0, \ \ \ \ \ \  r \gg 1.
   \label{5AHOas}
 \ee
This equation obviously possesses just the non-localized,
asymptotically oscillatory solutions (remember that $\mu$ is now
purely imaginary). As a consequence, all the ``bound-state"
solutions obtained in \cite{cinani} remain non-localized and
correspond in fact to the scattering states. In the other words,
the implicit formula (24) of \cite{cinani} for the energies $E_n$
is incorrect while the respective definitions (25) and (26) of the
related Dirac's ``bound-state" wave functions remain
unnormalizable and, hence, irrelevant.

An explanation of such a failure of the method returns us to the
transition between formulae  (11) and (12) in \cite{cinani} where
the latter formula has been erroneously declared to be the only
solution compatible with the correct asymptotic bound-state
boundary conditions. In fact, such an argument fails completely
when $\mu$ becomes imaginary.

(ii) Let us now move to the second, pseudospin-symmetric regime
where the authors of ref.~\cite{cinani} claim the existence of the
bound-state solutions of the Dirac bound-state problem
(\ref{6AHO}) in the two separate subcases again [let us mark them
as (ii.a) and (ii.b) in what follows]. We arrived at a different
conclusion.

(ii.a) In eq.~(\ref{6AHO}) let us follow ref.~\cite{cinani} and
assume, firstly, that $\mu^2_{(ii)}=\mu^2<0$ while
$\nu^2_{(ii)}=\nu^2>0$ and $\nu^2-\mu^2 \equiv \gamma^2>0$. This
leads to the problem
 \be
 -{a^2}\,\frac{d^2}{dr^2}\,G
 (r) +{\mu^2_{(ii)}}
 \,G
 (r)=0, \ \ \ \ \ \  r \gg 1
   \label{6AHOas}
 \ee
which parallels eq.~(\ref{5AHOas}). Hence, our conclusions remain
the same. The implicit formula (34) for the bound-state energies
$E_n$ in \cite{cinani} is incorrect while the related wave
functions (35) and (36) are not normalizable and represent merely
a random sample of the scattering states.

(ii.b) In the last subcase, equation~(\ref{6AHO}) with the
alternative choice of the real $\mu$ (with positive square
$\mu^2_{(ii)}=\mu^2\geq 0$) is to be combined with the purely
imaginary $\nu$ such that $\nu^2_{(ii)}=\nu^2\leq 0$ while
$\mu^2-\nu^2 \equiv \delta^2\geq 0$. Once we try to re-analyze
such a model in present setting, we simply imagine that we deal
with the Schr\"{o}dinger-like equation
 \be
 \left (
 -\frac{d^2}{dr^2} +\kappa^2 [M+E-\Delta(r)]
 \right )
 \,G_n
 (r)=0
   \label{AHOq}
 \ee
where $\kappa^2=M-E+C>0$ and $M+E>0$ in the original notation.
There are no bound states for the positive $\Delta_0>0$, of
course. This observation is not incompatible with the observation
made in ref. \cite{cinani} where the wave functions (39) and (40)
with the most elementary energy formula (38) represented just the
single state lying at the very boundary $\mu=0$ of the domain,
anyhow. Nevertheless, a return to the latter $\mu=0$ solution of
eq.~(\ref{6AHO}) ``on the boundary" reveals that it does not need
to exist at all. For example, we may verify that the insertion of
the parameters $\mu=\nu=0$ in the general solution (say, of the
form similar to eq.~(11) in \cite{cinani}) leads to the { exact}
wave function which is a constant. This constant must vanish due
to the boundary conditions.

In our final remark we have to emphasize that the situation in the
pseudospin-symmetric regime is by far not so hopeless as it seems
to be. As long as our potential functions are always
asymptotically vanishing, our last equation (\ref{AHOq}) {\em
will} possess (a finite number of) bound states whenever the
effective potential well proves, paradoxically, sufficiently
strongly repulsive, $-\Delta(0) = +\Delta_0 \leq
\Delta_0^{(critical)} < 0$.

The existence of the latter family of bound states was not
considered in ref.~\cite{cinani} at all. We repeat that in a
search for a new relativistic Woods-Saxon model, one has to opt
for a ``paradoxical" choice of the repulsive barrier with
$\Delta_0<0$ (i.e., with $\Delta(r)> 0$ at all $r$). This choice,
obviously, represents {\em the only eligible} pseudospin-symmetric
model where we would be allowed to construct bound states. The
details of such a construction are entirely straightforward and
may be left to the readers.

\section*{Acknowledgment}

Work partially supported from IRP AV0Z10480505 and by GA AS in
Prague, grant No.~A 1048302.

\end{document}